\documentclass{iau}
\usepackage{graphicx}

\title[$\eta$ Aql: pulsation and magnetic field] 
{Spectropolarimetric study of the classical Cepheid $\eta$ Aql:\\ pulsation and magnetic field}

\author[Butkovskaya et al.]   
{V.~Butkovskaya$^1$, S.~Plachinda$^1$, D.~Baklanova$^1$
 \and V.~Butkovskyi$^2$}

\affiliation{$^1$Crimean Astrophysical Observatory of Taras Shevchenko National University of Kyiv, \\ 98409,  Nauchny, Crimea, Ukraine,\\ email: {\tt varya@crao.crimea.ua} \\[\affilskip]
$^2$Taurida National V.\,I.\,Vernadsky University, \\ 95007, Vernadskogo str. 4, Simferopol, Crimea, Ukraine}

\pubyear{2014}
\volume{301}  
\pagerange{1--2}
\jname{Precision Asteroseismology}
\editors{J.A. Guzik, W.J. Chaplin, G. Handler \& A. Pigulski, eds.}
\begin{document}

\maketitle

\begin{abstract}
We report the results of spectropolarimetric study of the classical Cepheid $\eta$ Aql. We found that the longitudinal magnetic field of $\eta$ Aql sinusoidally varies with the radial pulsation period, while the amplitude $B$, mean field $B_0$, and phases of maximum and minimum field change from year to year. We hypothesize that possible reasons of those variations are stellar axial rotation or dynamo mechanisms.
\end{abstract}

\firstsection 
\section{Introduction}
Currently, the question of pulsation modulation of magnetic field in stars, both with convective and radiative envelopes, is still opened. 

\textbf{RR Lyr.} 
\cite[Babcock (1958)]{bb58}
reported a detection of a magnetic field in RR~Lyr. The longitudinal component of the field was found to be variable from $-$1580 to +540 G, but showed no correlation with the pulsation cycle of the star. 
\cite[Romanov et al.~(1987, 1994)]{romanov87}
also registered significant magnetic field in RR Lyr, and found the field to be variable  with an amplitude of up to 1.5 kG over the pulsation cycle. On the other hand,
\cite[Preston (1967)]{prest67}
and
\cite[Chadid et al.~(2004)]{chadid04}
detected no convincing evidence of a photospheric magnetic field in the star in the years 1963--1964 and 1999--2002, respectively.

\textbf{$\eta$ Aql.} Photoelectric magnetometer observations of $\eta$ Aql, performed by 
\cite[Borra et al.~(1981, 1984)]{borra81,borra84}, 
detected no magnetic field in this star. 
\cite[Plachinda (2000)]{[plach00}
was the first who detected magnetic field on $\eta$ Aql and reported pulsation modulation of the longitudinal component from $-$100 to +50 G. 
\cite[Wade et al.~(2002)]{wade02} 
detected no statistically significant longitudinal magnetic field in $\eta$ Aql during 3 nights in 2001 and concluded that $\eta$ Aql is a non-magnetic star, at least at a level of 10 G. 
\cite[Grunhut et al.~(2010)]{grunhut04} 
registered clear Zeeman signatures in Stokes V parameter for $\eta$~Aql and eight other supergiants. 

\textbf{$\gamma$ Peg.} 
\cite[Butkovskaya \& Plachinda (2007)]{butk07} 
reported the modulation of the longitudinal magnetic field in $\beta$ Cephei-type star $\gamma$ Peg (B2 IV)~with the amplitude of about 7 G over the 0.15-day pulsation period of the star.

In order to shed some light on the problem of pulsational modulation of stellar magnetic fields, we continued spectropolarimetric monitoring of $\eta$ Aql during 60 nights between 2002 and 2012 using Coud\'e spectrograph at the 2.6-m Shajn telescope of the Crimean Astrophysical Observatory (Ukraine). 

\section{Results}
\begin{figure}[!ht]
\begin{center}
 \includegraphics[width=4.0in]{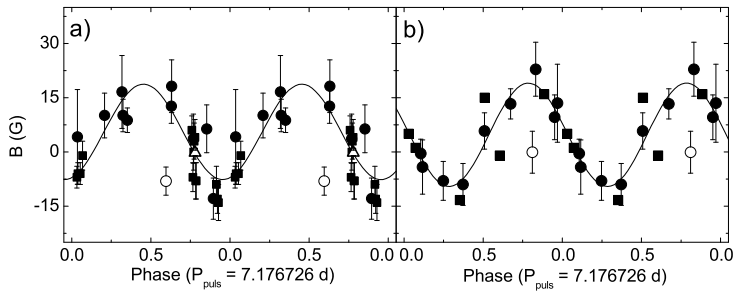} 
 \caption{Longitudinal magnetic field of $\eta$ Aql folded with the 7.176726-day pulsation period: a) our data obtained in 2002 (\textit{closed and open circles}), data by 
\cite[Wade et al.~(2002)]{wade02} 
(\textit{black squares}),
and
\cite[Grunhut et al.~(2010)]{grunhut04} 
(\textit{open triangles});
b) our data obtained in 2004 (\textit{closed and open circles}), data by 
\cite[Borra et al.~(1981, 1984)]{borra81,borra84} (\textit{black squares}). 
Fitting sinusoids are shown by strong lines. Open circles represent our data that have not been taken into account in the fits.}
   \label{fig1}
\end{center}
\end{figure}

The technique of Zeeman splitting, used for the measurement of the longitudinal magnetic field, is described in detail by 
\cite[Butkovskaya \& Plachinda (2007)]{butk07}.
We folded all values of the measured longitudinal magnetic field with the pulsational period according to the pulsation ephemeris: JD $= 2450100.861+7.176726E$ \cite[(Kiss \& Vink\'o 2000)]{kiss00}, where $E$ is the number of pulsation cycles. As an example, the pulsation modulation of the longitudinal magnetic field of $\eta$ Aql in 2002 and 2004 is illustrated in Fig.
\ref{fig1}.
We found that the magnetic field sinusoidally varies in phase of the radial pulsation of $\eta$ Aql. This confirms the previous conclusions of \cite[Plachinda (2000)]{plach00}. However, the amplitude $B$, mean field $B_0$, and phases of maximum and minimum field are changing from year to year (see Table 1 where the F-test indicates the statistical reliability of the detected variability for each year). The possible reason for those variations is stellar axial rotation or dynamo mechanisms. 
\begin{table}
  \begin{center}
  \caption{Parameters of the variability of the longitudinal magnetic field of $\eta$~Aql.}
  \label{tab1}
 {\scriptsize
  \begin{tabular}{|c|c|c|r|r|r|}\hline 
{\bf Year} & {\bf Phase max} & {\bf Phase min} & {\bf Amplitude $B$, G} & {\bf Mean $B_{0}$, G} & {F-test}\\ 
 \hline
2002 & 0.45 & 0.95& 12.7 $\pm$ 2.2 & 5.5 $\pm$ 1.5 & 0.99 \\ 
2004 & 0.78 & 0.28& 13.9 $\pm$ 2.4 & 4.7 $\pm$ 1.5 & 0.99 \\ 
2010 & 0.38 & 0.88&  4.3 $\pm$ 1.1 &$-$3.1 $\pm$ 1.1 & 0.98 \\ 
2012 & 0.60 & 0.10&  4.2 $\pm$ 2.3 &$-$0.7 $\pm$ 1.5 & 0.69 \\ \hline
  \end{tabular}
  }
 \end{center}  
\end{table}

\end{document}